%% file: main.tex
\documentclass[final]{IEEEtran}

\usepackage{amsmath}
\usepackage{graphicx}
\usepackage{acronym}
\usepackage[colorlinks=true, allcolors=blue]{hyperref}
\usepackage[font=footnotesize]{caption}
\usepackage[font=footnotesize]{subcaption}
\usepackage[dvipsnames]{xcolor}
\definecolor{mygray}{gray}{0.6}

\usepackage{tikz}
\usetikzlibrary{calc}
\input{acs}
\title{Leveraging RIS-Enabled Smart Signal Propagation for Solving Infeasible Localization Problems}
\author{\IEEEauthorblockN{Kamran Keykhosravi, Beno\^{i}t Denis, George C.~Alexandropoulos, \\Zhongxia~Simon~He, Antonio Albanese, Vincenzo Sciancalepore, and Henk Wymeersch}

}

\begin{document}
\maketitle

\begin{abstract}

Reconfigurable intelligent surfaces (RISs) have tremendous potential for both communication and localization. While communication benefits are now well-understood, the breakthrough nature of the technology may well lie in its capability to provide location estimates when conventional approaches fail, (e.g., due to insufficient available infrastructure). A limited number of example scenarios have been identified, but an overview of possible RIS-enabled localization scenarios is still missing from the literature. In this article, we present such an overview and extend localization to include even user orientation or velocity. In particular, we consider localization scenarios with various numbers of RISs, single- or multi-antenna base stations, narrowband or wideband transmissions, and near- and far-field operation. Furthermore, we provide a short description of the general RIS operation together with radio localization fundamentals, experimental validation of a localization scheme with two RISs, as well as key research directions and open challenges specific to RIS-enabled localization and sensing.
\end{abstract}

\section{Introduction}

Radio localization via wireless network infrastructure is a service stemmed from governmental mandates on positioning emergency calls by network operators. Over time, especially after introducing dedicated signals and procedures in 3GPP R16 \cite{3gppR16}, radio localization found many other applications, including navigation, network optimization, geo-targeting, and augmented reality~\cite{del2017survey}, particularly for scenarios where the Global Positioning System (GPS) is insufficient (or not available), such as indoor environments, urban canyons, and tunnels.  With cellular localization, the \ac{ue} state (comprising \ac{ue} location, time bias, but possibly also the orientation, velocity, etc.) can be estimated based on a variety of measurements from the received signal, including the signal strength, \ac{toa}, \ac{aoa}, and \ac{aod}. In the fifth generation (5G) of wireless systems, radio localization can be accurately performed thanks to a multitude of antennas and large radio bandwidth. In the future sixth generation (6G), radio localization is envisioned to be even more ubiquitous as well as cost- and energy-efficient by utilizing \acp{ris}~\cite{RISE6G_COMMAG}.

\acp{ris} are thin surfaces, comprising many small unit cells, each of which is capable of controlled modulation of the impinging signal's phase and amplitude before scattering it back into the environment \cite{LIS_twc2018,risTUTORIAL2020}. The response of an \ac{ris} can be optimized to maximize the \ac{qos} of the wireless operation. Comparing with provisioning an additional multi-antenna \ac{bs} or relay, \acp{ris} are a much more cost- and energy-efficient alternative for providing ubiquitous and high-capacity connectivity~\cite{LIS_twc2018}. Due to the very limited power consumption, they are even suitable for delivering 3D ubiquitous features being installed on board of \acp{uav} in emergency scenarios~\cite{alb_commag21}. Additionally, they can be easily installed on common surfaces such as walls and billboards, they can contain a vast  number of unit cells, and are either passive or have very few active elements. 
 Like many other technologies that are introduced to enhance communication systems, \acp{ris} can be repurposed for  radio localization and can boost or, more notably, enable localization by providing a strong indirect signal path as well as an additional localization reference~\cite{Wymeersch_2020}. 

In this magazine paper, we aim to explore scenarios where localization (or, in general, estimation of \ac{ue}-state parameters) is enabled by \acp{ris}. These scenarios include \ac{mimo}, \ac{miso}, \ac{simo}, and \ac{siso} setups; wideband or narrowband signalling; and far- or near-field regimes. These \ac{ris}-enabled scenarios rely on extremely limited  \ac{bs} infrastructure, thereby leading to unexpected cost and energy savings and they are therefore important to be widely known in the community. We discuss how geometrical information obtained from the received signal via RISs can enable localization and sensing in each scenario. We also present an experimental localization example at $60$ GHz with one \ac{bs} and two \acp{ris} to support our discussion.

\begin{figure}
    \centering
    \begin{subfigure}[b]{0.45\textwidth}
         \centering
         \includegraphics[width=\textwidth]{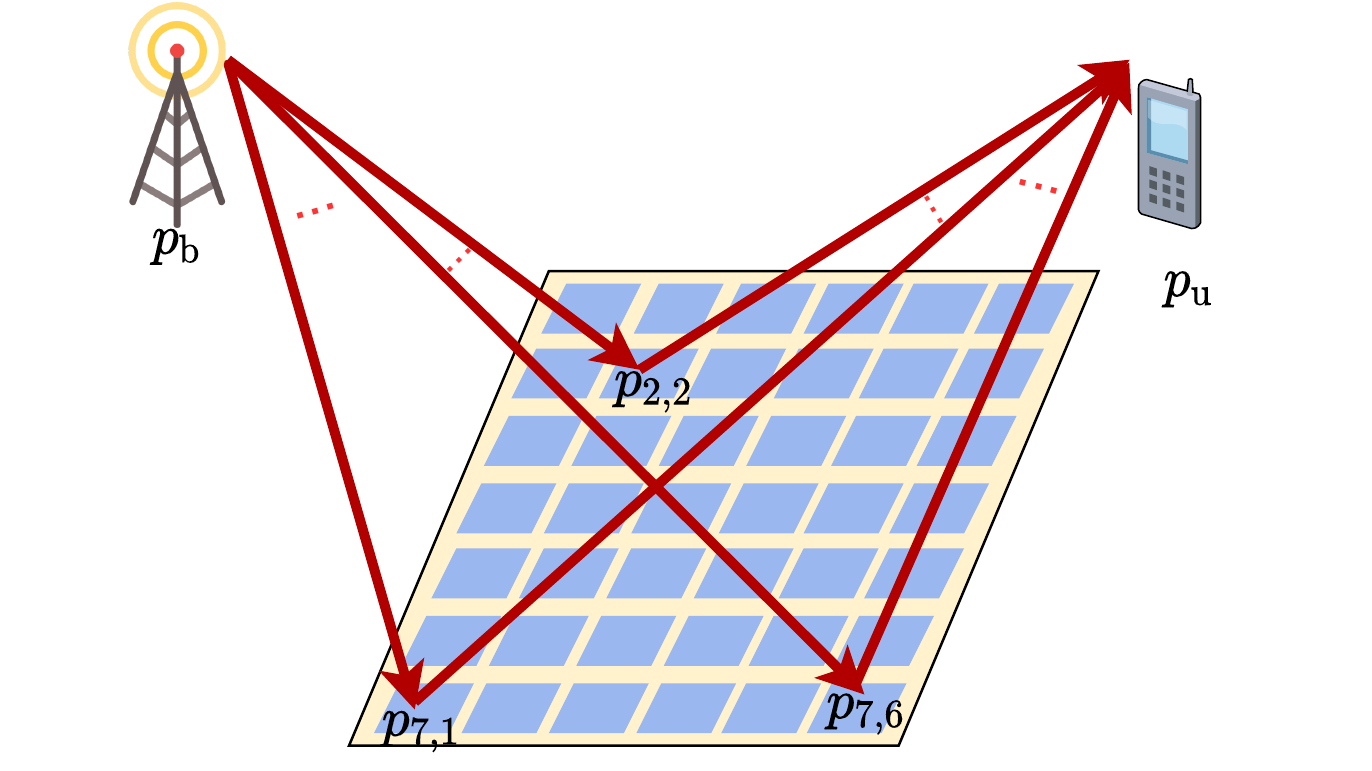}
         \caption{Near-field regime.}
         \label{fig:nf}
     \end{subfigure}
         \begin{subfigure}[b]{0.45\textwidth}
         \centering
         \includegraphics[width=\textwidth]{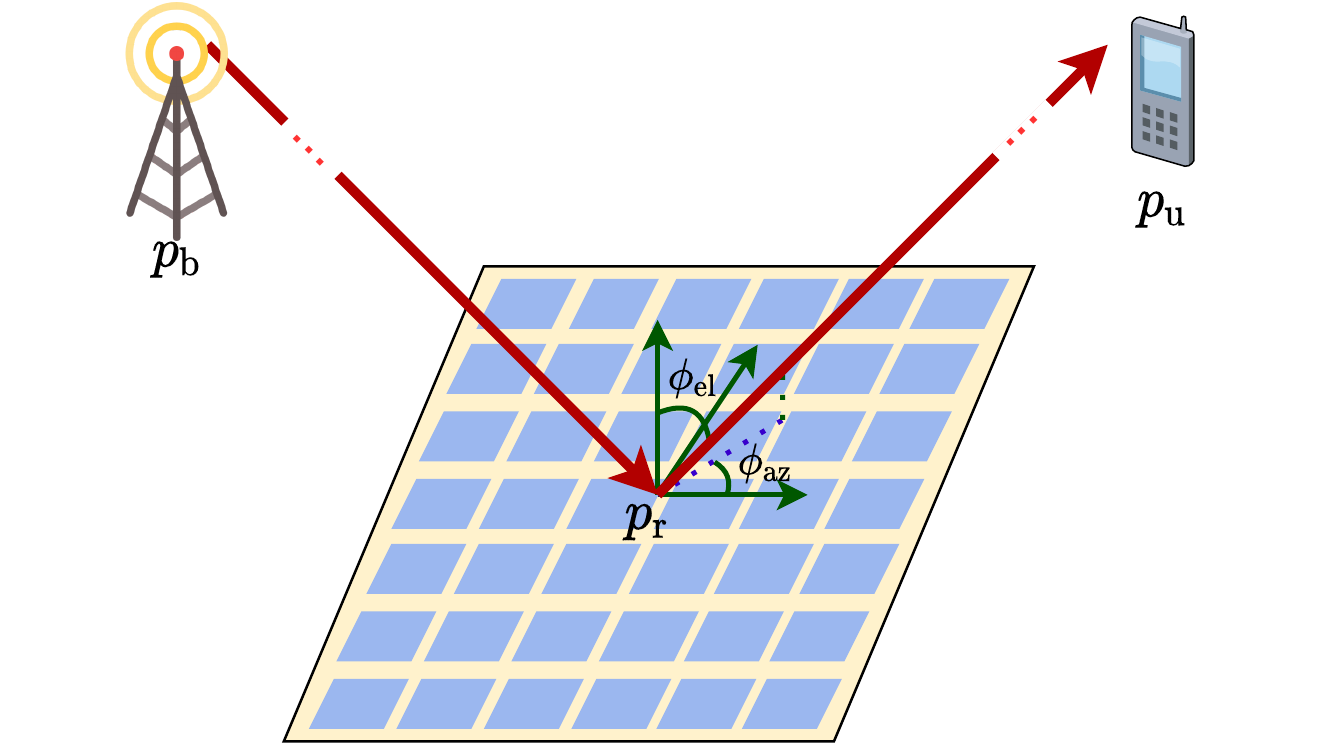}
         \caption{Far-field regime.}
         \label{fig:ff}
     \end{subfigure}
     \caption{The near-field (a) and far-field (b) regimes of the RIS-enabled signal propagation with respect to the RIS. The received signal in the downlink can be expressed as function of the UE position in (a) and of the signal's \ac{aod} in (b).
     }
    \label{fig:NFFF}
\end{figure}

\begin{figure*}
    \centering
    \includegraphics[width=\linewidth]{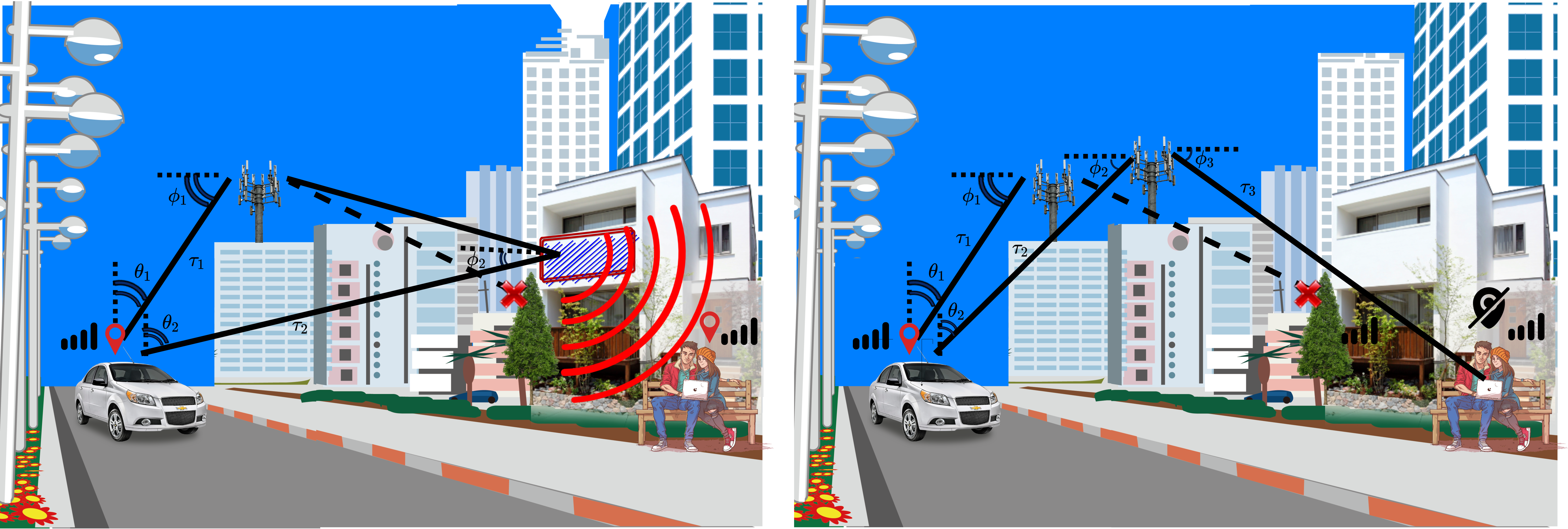}
    \caption{RIS-enabled localization: in the left, localization of the laptop residing the near field of an RIS is feasible with a single BS, while it is unfeasible when placed on the far field of two BSs. The symbols $\tau$, $\phi$, $\theta$ are used to indicate \acp{toa}, \acp{aod}, and  AoAs, respectively. }
    \label{fig:vision}
\end{figure*}

\section{RIS operation mode to support localization}
A planar array of ultra-thin elementary electronic circuits or metameterials, each capable of realizing distinct electro-magnetic (EM) states, is known as RIS \cite{LIS_twc2018}. Its dynamic reconfigurability according to desired wireless networking objectives is enabled by multitudes of ultra-low power unit elements, such as positive intrinsic negative (PIN) diodes or varactors. For example, by appropriately configuring the ON/OFF state of the PIN diodes or the bias voltage of the varactors, the resulting macroscopic transformations of the radio waves impinging at the RIS can be controlled, offering desired reflective beamforming towards intended receivers.

For quasi-free-space beam manipulation, which concerns the main application scenario for RISs in wireless communications~\cite{Tang_2021}, a fine-grained control over the reflected EM field is essential for accurate reflective beamforming. This fact motivated researchers to rely on RIS elements of sub-wavelength size (e.g., $\lambda/10$ with $\lambda$ being the wavelength,
despite inevitable strong inter-element mutual coupling, and well-defined, possibly gray-scale-tunable, EM properties. In contrast, in rich scattering environments, the wave energy is statistically equally spread throughout the wireless medium and the ensuing ray chaos implies that rays impact the RIS from all possible, rather than one well-defined, direction. Instead of creating a directive reflective beam, the goal becomes the manipulation of as many ray paths as possible. This manipulation may either aim at tailoring those rays to create constructive interference at a target location or to efficiently stir the field \cite{alexandg_2021}. These manipulations can be efficiently realized with half-wavelength-sized elements, which enable the control of more rays with a fixed amount of electronic components, as compared to RISs equipped with their sub-wavelength counterparts. In terms of positioning for both aforementioned wireless conditions, RISs offer additional degrees of freedom that can be leveraged for tackling, otherwise infeasible, localization problems. Up to date, in rich-scattering media, EM wave fingerprinting in conjunction with supervised learning approaches are mainly deployed \cite{alexandg_2021}, whereas multipath channels with a dominant \ac{los} component constitute the largely considered application conditions, which are the subject of this article.


The channel modeling of RIS-enabled smart wireless environments is an active area of research \cite{tang2020wireless}. 
For localization purposes, sparse parametric models are often used, where the channel is represented via a few geometric components. Figure\,\ref{fig:nf} illustrates an one-ray SISO system including one RIS, where both the BS and UE are in the near field of the RIS. In this case, the received signal in the downlink can be calculated as a sum of individual rays reflected from each RIS element at the UE location. This signal is a function of the transmitted symbols, the BS position $\mathbf{p}_{\text{b}}$, the positions of the RIS elements  $\mathbf{p}_{i,j}$, and the UE position  $\mathbf{p}_{\text{u}}$, among which only the last one is unknown and needs be estimated. In Fig.\,\ref{fig:ff}, we demonstrate the same SISO setup but with the BS and UE being in the far field of the RIS, where the distances between the UE and BS is much larger than the RIS size. In this case, the wireless channel can be described by the AoD, which consists of elevation and azimuth factors (represented by $\phi_{\text{el}}$ and $\phi_{\text{az}}$ in Fig.\,\ref{fig:ff}).

In general, localization requires more network infrastructure than establishing a wireless communication link, since information from multiple signal paths is required to estimate a \ac{ue} position. In Fig.\,\ref{fig:vision}, we compare two wireless systems with two UEs (a car and a laptop): the one (at left) includes a BS and an RIS and the other (at right) two BSs. The position of the car can be estimated in both scenarios from the angle and delay measurements from the BS(s) and/or the RIS. However, the location of the laptop can only be estimated in the left scenario, where it is placed in the near-field region of the RIS. In this case, the wavefront curvature of the received signal can be used for localizaton. In the right scenario, the laptop only receives signals from the BS at the far-field field, which would be enough to establish a communication link, but not sufficient for inferring its position.

 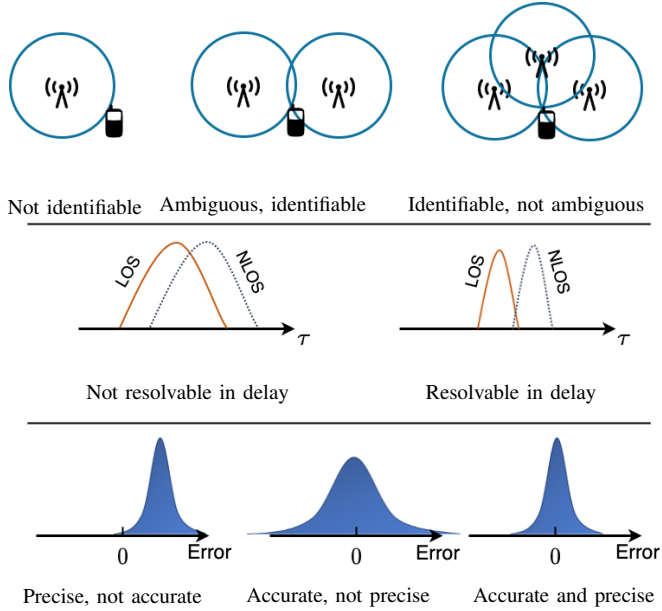
\begin{figure}
    \centering
    \begin{tikzpicture}
    \node(image) [anchor=south west] at (0,0)  { \includegraphics[width=\linewidth]{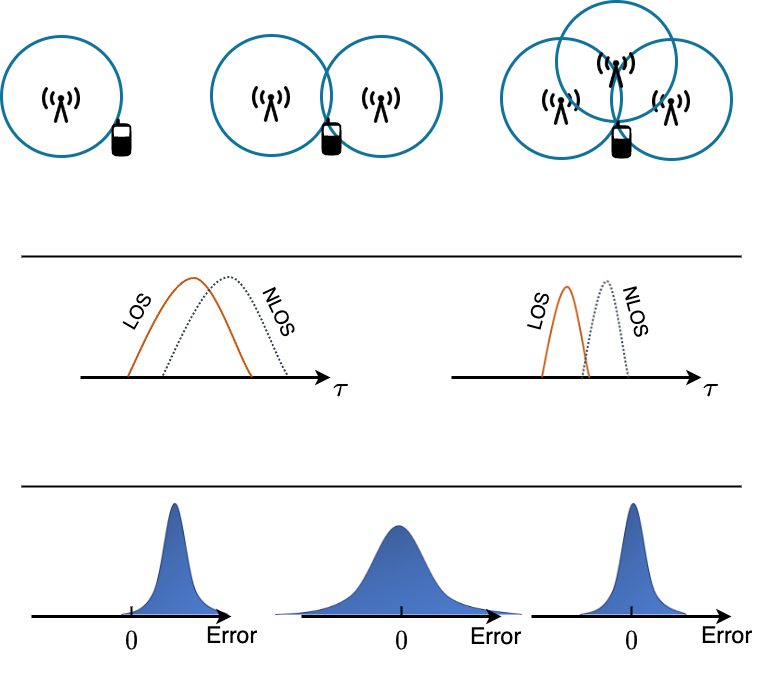}};
    \node at (1,5.2){\footnotesize Not identifiable};
    \node at (3.5,5.2){\footnotesize Ambiguous, identifiable};
    \node at (7,5.2){\footnotesize Identifiable, not ambiguous};
    \node at (1.5,0){\footnotesize Precise, not accurate};
    \node at (4.5,0){\footnotesize Accurate, not precise};
    \node at (7.5,0){\footnotesize Accurate and precise};
    \node at (2.5,2.7){\footnotesize Not resolvable in delay};
    \node at (6.8,2.7){\footnotesize Resolvable in delay};
    \end{tikzpicture}
    \caption{Identifiability, ambiguity, resolvability, precision, and accuracy in 2D localization problems. }
    \label{fig:iden}
\end{figure}

\section{Wireless localization fundamentals}\label{sec:localization}
Radio localization methods are based on the premise that received waves convey geometric information about the propagation channel, which can be in turn used to determine the locations of the wirelessly connected devices (i.e., \acp{ue}). Broadly speaking, radio localization methods can be categorized as \emph{data-driven} and \emph{model-driven}. In the former category, we have methods such as fingerprinting and artificial-intelligence-assisted localization, which rely on rich features of the received radio signal, but without a structure that is easy to model. The latter category of model-driven methods harnesses approximate statistical relations between the received radio signal and the signal propagation geometry (including the \ac{ue} location), and form the bulk of radio localization techniques used in practice. Model-based methods have a large number of benefits over data-driven methods. They rely on decades of signal processing methods and optimization techniques, usually offering lower complexity than data-driven approaches, and are accompanied with performance bounds that provide strong guarantees on optimality and reliability.

Practical radio-based localization methods in 5G systems rely on time-based measurements with respect to several \acp{bs}, where the propagation time of the direct \ac{los} signal path to or from a \ac{bs} is proportional to the distance between the \ac{ue} and \ac{bs}, but also includes an unknown clock bias of the \ac{ue} with respect to the \ac{bs}. As a specific example, under \ac{tdoa}, measurements with respect to $4$ synchronized \acp{bs} are needed to {solve the 3D positioning and 1D clock bias estimation problem out of one-way transmissions (i.e., uplink or downlink)}. In contrast, under two-way \ac{rtt} measurements, when signal exchanges between BS and UE naturally remove the timing uncertainty, $3$ \acp{bs} are sufficient for \ac{ue} localization. To complement delay-based measurements, angle measurements (\ac{aoa}  and \ac{aod}) are  employed, which, based on a codebook of directional beams, constrain the \ac{ue} to lie within an angular sector. The sector size depends on the beamwidth, with the latter being inversely proportional to the array size.   

The performance of  model-based localization methods depends on several fundamental concepts: \emph{identifiability}, \emph{ambiguity}, \emph{resolution}, \emph{precision}, and \emph{accuracy}. These are visualized in Fig.~\ref{fig:iden}. 
Note that this does not correspond to an exhaustive list of metrics (which would include latency, update rate, integrity, etc.), but merely to convey the fundamentals.

\begin{figure*}
    \centering
    \includegraphics[width=\linewidth]{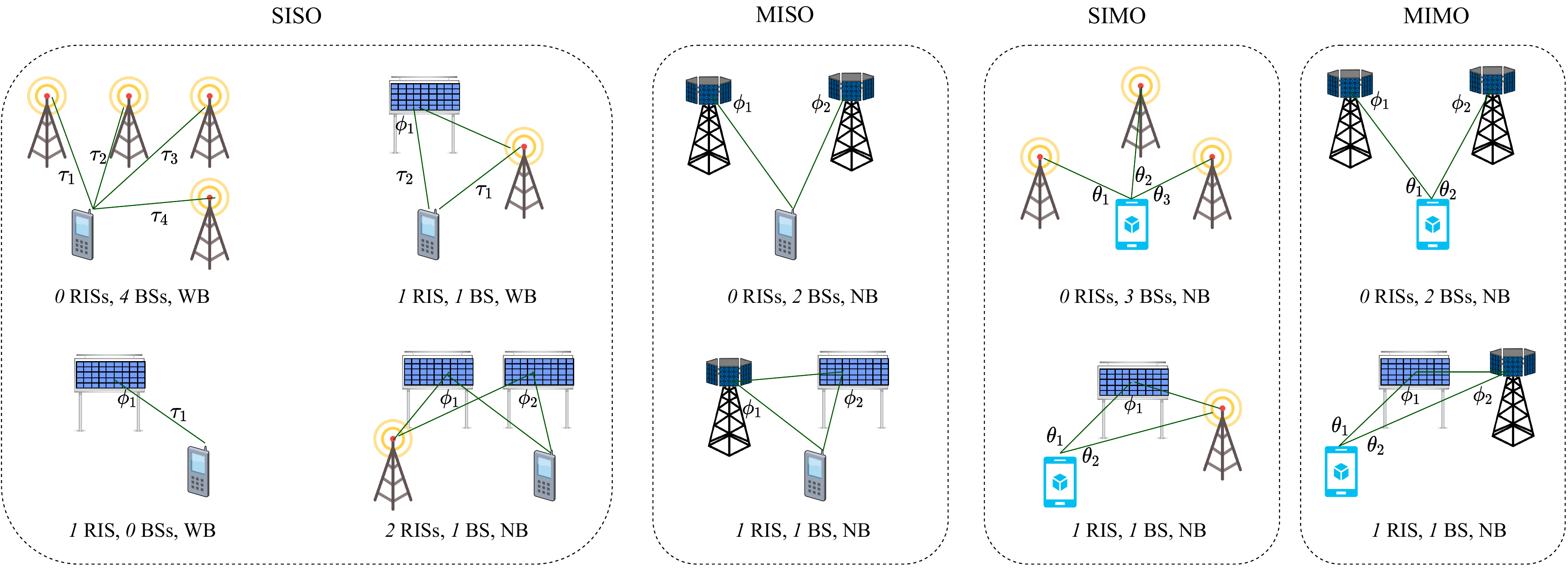}
    \caption{Scenarios of feasible 3D \ac{ue} positioning. The symbols $\tau$, $\phi$, and $\theta$ indicate \acp{toa}, \acp{aod}, and  AoAs, respectively. }
    \label{fig:scenarios}
\end{figure*}

\begin{itemize}
    \item \textbf{Precision and accuracy} are shown in the bottom part of Fig.~\ref{fig:iden}, and refer, respectively, to the spread and bias of localization errors, but are usually treated jointly (with either term being commonly used, though we will use accuracy herein) through the localization error statistics (e.g., mean or median error, 90\% confidence interval). Bounds on the achievable accuracy include the Cram\'{e}r-Rao bound, the Weiss-Weinstein bounds, and Ziv-Zaka\"{i} bounds. In principle, the accuracy is not bounded from below, and can be arbitrarily improved by increasing the \ac{snr}. In addition to link-level \ac{snr}, accuracy is also determined by the deployment geometry (i.e., \ac{ue} relative position with respect to BSs), also referred to as \ac{gdop} (e.g., in the GPS community). Accuracy is arguably the  definitive metric for any localization method, as it accounts for both resolution and identifiability. 
\item \textbf{Resolution} in our context refers to the separability of perfectly correlated radio propagation paths in at least one domain (which could be delay/Doppler, as commonly used in radar, but also angle or polarization). An example of delay resolution is shown in the middle part of Fig.~\ref{fig:iden}. Signal paths that are not resolvable will be interpreted as a single path, thus limiting the accuracy (irrespective of the \ac{snr}), and leading to worse performance than predicted by analytical bounds. The resolution is bounded by the available physical resources, i.e., bandwidth for delay/distance resolution (and conversely time for frequency/Doppler resolution), and antenna array aperture for angle resolution. Accordingly, aiming at achieving better resolution (rather than better accuracy), more bandwidth and larger arrays are, in general, required to improve localization. This explains the poor localization performance of 4G and earlier generations, which relied on limited bandwidths and small arrays. 5G and 6G provide a leap forward in delay and angle resolution through the combination of larger bandwidths at high carriers and integrated massive multi-antenna systems \cite{bartoletti20185g}. 
\item \textbf{Ambiguity and non-identifiability} may still occur, 
even with very high resolution. In particular, the solution to the localization problem may be  ambiguous (i.e., there are several distinct solutions, as shown in the top center part of Fig.~\ref{fig:iden}) or the problem may not be identifiable (i.e., there exists a continuous space of solutions, as shown in the top left part of Fig.~\ref{fig:iden}). This happens, for instance, when the \ac{los} signal from one BS is missing due to an obstruction or when the infrastructure deployment/coverage is not sufficient. Ambiguity can generally be resolved by prior information (e.g., an ambiguous GPS solution far away from the earth's surface can readily be discarded) and occur generally intermittently. On the other hand, identifiability is more harmful, as there are many equally valid solutions to the localization problem, most of which cannot be discarded based on external information. We note that identifiability is related to, but distinct from observability in control theory, where observability refers to the ability to estimate the user location over time. 

\end{itemize}

There are several (equivalent) approaches to assess identifiability, including geometry and Fisher information analysis. The geometric approach is generative, in the sense that, it constructs the manifold that constrains the solution based on each measurement, and then, determines the dimensionality of the intersection of these manifolds. An identifiable problem would thus have a zero-dimensional manifold as solution. Instead, a Fisher information analysis {aims at characterizing the amount of location information conveyed by measurements, given both the known statistics of the latter and the true UE location. More specifically, it} determines the local curvature of the likelihood function. This curvature is described by a matrix, called the Fisher information matrix. When this matrix is not full-rank, the likelihood is locally flat along at least one dimension in the vicinity of the true location, rendering the problem infeasible. An example of an infeasible localization problem is \ac{tdoa}-based localization with $2$ \acp{bs} (the intersection of two hyperboloids is an 1D manifold). 

In the following section, we discuss the expected advantages of \ac{ris}-empowered settings in terms of localization feasibility, when compared to conventional settings in similar scenarios and operating contexts.

\section{RIS-enabled localization scenarios}

\begin{table*}
\centering
\caption{Identifiability analysis of 3D \ac{ue} positioning in the downlink. The abbreviations pos, vel, and ori stand for position, velocity, and orientation, respectively.}\label{table:overview}
\resizebox{0.99\textwidth}{!} {
\begin{tabular}{|c|c|c|c|c|c|}
\hline 
 & \textbf{Scenario} &\textbf{Signalling} & \textbf{Measurements} & \textbf{Identifiable State} &  \textbf{Positioning is possible also}\tabularnewline
\hline 
\hline 
\textbf{SISO}& $0$ RISs, $4$ BSs &  WB & TDoA & 3D pos, clock, 3D vel & with $3$ BSs and RTT measurements  \tabularnewline
\hline 
\textbf{SISO}& $1$ RIS, $1$ BS &  WB &  TDoA, AoD & 3D pos, clock, 2D vel & in near-field  w/o LOS to BS\tabularnewline
\hline 
\textbf{SISO}& $2$ RISs, $1$ BS &  NB &   AoD & 3D pos, 3D vel &  w/o LOS to BS\tabularnewline
\hline 
\textbf{SISO}& $1$ RIS, $0$ BSs &  WB &  RTT, AoD & 3D pos, 1D vel & N/A\tabularnewline
\hline
\textbf{MISO} & $0$ RISs, $2$ BSs &  NB & AoD & 3D pos, 2D vel & N/A\tabularnewline
\hline 
\textbf{MISO} & $1$ RIS, $1$ BS &  NB & AoD & 3D pos, 2D vel & in near-field  w/o LOS to BS\tabularnewline
\hline 
\textbf{SIMO} & $0$ RISs, $3$ BSs & NB &   AoA  & 3D pos, 3D vel, 3D ori & N/A\tabularnewline
\hline 
\textbf{SIMO} & $1$ RIS, $1$ BS & NB &  AoD, AoA & 3D pos, 2D vel, 3D ori & N/A\tabularnewline
\hline 
\textbf{MIMO} & $0$ RISs, $2$ BSs & NB &  AoD, AoA  & 3D pos, 2D vel, 3D ori & N/A\tabularnewline
\hline 
\textbf{MIMO} & $1$ RIS, $1$ BS & NB &   AoD, AoA & 3D pos, 2D vel, 3D ori & in near-field  w/o LOS to BS\tabularnewline
\hline 
\end{tabular}
}
\end{table*}

In this section, we present a number of localization scenarios wherein \ac{ue} 3D positioning can be achieved. In addition to positioning, we discuss whether the \ac{ue} velocity, clock bias, and orientation are identifiable. We focus on the downlink direction since it facilitates the positioning of multiple \acp{ue} at once. We consider minimal scenarios, meaning that if a single BS or RIS is removed, 3D positioning is no longer possible. Besides, we assume that all the BSs are synchronized with each other, while the UE is not. 

In terms of signalling, we consider both \ac{nb} and \ac{wb} signals, where only in the latter case the \ac{toa} estimation can be performed.  Moreover, depending on the scenario, the \acp{bs} and the \ac{ue} are equipped with either a single or multiple antennas, where angle measurements are only possible in the latter case. We list relevant scenarios in Table~\ref{table:overview}, which are also illustrated in Fig.~\ref{fig:scenarios}.

\subsection{SISO localization}

\textbf{ SISO with $\mathbf{4}$ BSs in the absence of an RIS.}\label{sec:scenario:Siso0Ris4BS}
We consider the standard localization protocol for cellular networks and assume that four synchronized BSs transmit \ac{wb} pilot signals to the UE, generating three \acp{tdoa} or four \acp{toa}. The UE position can be estimated by calculating the intersection of the three hyperboloids corresponding to the three \ac{tdoa} measurements. Next, based on the UE position estimate and the measured \acp{toa}, we can obtain the clock bias allowing the UE to be synchronized to the \acp{bs}. By measuring the RTTs instead of the \acp{tdoa} , UE localization becomes feasible even with three BSs via the intersection of the three spheres identified by the three RTTs and centered in the corresponding BSs. Lastly, we can derive the UE velocity vector in 3D by means of the four measured Doppler shifts (i.e., radial velocities).

\textbf{SISO with $\mathbf{1}$ RIS and $\mathbf{1}$ BS.}\label{sec:scenario:Siso1Ris1BS}
Let us assume a \ac{siso} system with a single RIS  as in \cite{keykhosravi2020siso}. By using \ac{wb} pilots, we can obtain the \acp{toa} for the direct (i.e., the path BS-UE) and the reflected (i.e., the path BS-RIS-UE) paths, from which we can calculate the resulting \ac{tdoa} and so the corresponding hyperboloid in 3D space. By using different RIS phase profiles at different transmission times, the \ac{aod} from the RIS to the UE can be estimated, which geometrically translates to a half-line. Therefore, we can calculate the UE position via the intersection between such half-line and the above-mentioned hyperboloid, while deriving the clock bias from the UE position estimate and the measured \acp{toa}. Moreover, we can exploit the Doppler shifts on the direct and the reflected paths to estimate the UE 2D velocity. Furthermore, if  the UE is in the near-field of the RIS, its position can be found by leveraging on the wavefront curvature, even if the direct path is blocked.  

\textbf{SISO with $\mathbf{2}$ RISs and $\mathbf{1}$ BS.}\label{sec:scenario:Siso2Ris1BS}
Let us assume a SISO system with two RISs. In this scenario, we can perform UE positioning even with  \ac{nb} signalling, which does not allow \ac{toa} estimation. Indeed, the UE position can be estimated via the intersection of the two half-lines corresponding to the \acp{aod} from the RISs. The direct BS-UE path does not carry any position information, thus localization can be performed even when the direct path is blocked. Nonetheless, the direct path provides Doppler information and enables us to estimate the UE velocity in 3D. In Section~\ref{sec:experiment}, we perform experimental measurements to further investigate this scenario. 

\textbf{SISO with $\mathbf{1}$ RIS in the absence of a BS.}\label{sec:scenario:Siso1Ris0BS}
A single-antenna full-duplex UE can estimate its own location by transmitting \ac{wb} pilots to the RIS and processing the reflected signals, thus (strictly) not requiring any BS (see \cite{keykhosravi2022ris}). Indeed, in this scenario, we can measure the RTT and the \ac{aod} from the RIS to the UE. Geometrically, they respectively correspond to a sphere centered in the RIS and a half-line originated in the RIS, whose intersection returns the UE position estimate. As the Doppler shifts can only be measured along the RIS-UE direction, the UE velocity can be estimated in 1D. However, if the UE motion direction is a-priori known, the 1D estimate fully identifies the UE velocity vector (e.g., for \acp{ue} moving along a highway).

\subsection{MISO localization}
We consider a MISO system, where multi-antenna BSs allow estimating the respective \ac{aod} towards the UE (see e.g., \cite{alessio_miso}). In this condition, we can perform UE positioning with only $2$ BSs and no RIS. Here, the UE position can be estimated by intersecting the two half-lines corresponding to the two \acp{aod} from the BSs. Since \acp{toa} measurements are not required, we can employ \ac{nb} pilots. In a similar fashion, UE localization can be achieved by replacing one of the BSs with an RIS, thus obtaining a MISO system with $1$ RIS and $1$ BS, and leveraging on the \acp{aod} from the BS and the RIS. 

\subsection{SIMO localization}\label{sec:scenario:SIMO}
In a SIMO system, it is possible to estimate the \acp{aoa} from the BSs at the UE in the UE's frame of reference, which depends on its orientation. We first consider a scenario with no RIS and $3$ BSs using \ac{nb} pilots. To show that the user position is identifiable, we define $\theta_{i,j}$ to be the angle between the direction of arrival from the $i$th BS and that of the $j$th one. Note that $\theta_{i,j}$ can be calculated based the measured AoAs and does not depend on the UE's orientation. Now consider any arbitrary plane that includes the $i$th and the $j$th BS. If the user is in this plane, then it should create a $\theta_{i,j}$ angle with these BSs. One can show that the locus of these points is described by an arc of a circle containing the two BSs and its reflection by the line that passes through the two BSs.
 Since this argument holds for all the planes that contain $i$th and $j$th BSs, to obtain all the points that  create $\theta_{i,j}$ angle with the two BSs, we should rotate this curve around the line that connects them. The generated surface is either the inner, or the outer part of a spindle Torus. Observe that the user should locate on the intersection of the three surfaces corresponding to $\theta_{1,2}$, $\theta_{1,3}$, and $\theta_{2,3}$, which, in general, has dimension zero (is a set of finite points). Therefore, the problem of user localization is identifiable. After estimating the UE position the UE orientation also can be found by means of any two \acp{aoa}.

Furthermore, localization is possible with $1$ RIS and $1$ BS. In this scenario, we can measure two \acp{aoa} and one \ac{aod} from the RIS.  Using the two AoAs, we can locate the user on (part of) a spindle Torus, whose intersection with the line corresponding the AoD locates the UE. Then the UE orientation can be estimated via the two AoAs.

\subsection{MIMO localization}
In MIMO systems, both \acp{aoa} and \acp{aod} can be estimated. Therefore, the UE localization is possible with no RIS and $2$ BSs or with $1$ RIS and $1$ BS. In both cases, the UE position can be estimated via the two \acp{aod} (by intersecting the two corresponding half-lines) while the UE orientation can be derived from the two AoAs. 

\subsection{Conventional vs RIS-enabled localization}
We hereinafter provide a qualitative comparison between two of the aforementioned scenarios: \emph{Scenario\,1: SISO with $2$ RISs and $1$ BS} and \emph{Scenario\,2: MISO with $0$ RISs and $2$ BSs}. Although both scenarios use two \acp{aod} to localize the UE, in the former case, the \acp{aod} are measured from the RISs, while in the latter one they are measured from the BSs. We compare the two methods in terms of cost, energy consumption and accuracy. RISs have lower manufacturing and installation costs than multi-antenna BSs since they are envisioned as cheap devices to be easily mounted on common surfaces, such as building facades, billboards, and walls. 
Furthermore, since RISs are almost passive devices, they consume much less energy than BSs. Therefore, \emph{Scenario\,1} is preferable to \emph{Scenario\,2} in terms of cost and power consumption. As far as localization accuracy goes, we need to consider two countering effects: number of antennas and cascaded path loss. As RISs can have many more reflecting elements than the BSs antennas, they can produce narrower beams and higher beam resolution. However, compared to the signal received directly from the BS, the reflected signal through an RIS suffers from a much larger path loss, which reduces the \ac{snr} and subsequently the localization accuracy. Therefore, the comparison in terms of localization accuracy depends on a number of factors, such as the RIS and BS array sizes, carrier frequency, and network geometry (e.g., BSs and RISs placement), which are tied to the specific parameters of the localization system.

\subsection{Localization in the presence of scattering}
In the previous discussion, all wireless links (i.e., BS--UE, BS--RIS, and RIS--UE) consist of only one LOS component. However, in practice, apart from the LOS path, many other signal components reach the UE after being scattered by the surrounding objects. If such components are not resolved at the UE, they interfere with the LOS signal at each link. This harms the estimation accuracy of the link's geometrical parameters and consequently deteriorates the localization accuracy.  

Nonetheless, if the NLOS components are resolvable in at least one domain (angle, delay, or Doppler), they can be separated from the LOS signal, but, more importantly, they can contribute to the UE localization and environmental mapping. In MIMO setups with no RISs, each NLOS path generated by a scatterer introduces three unknowns (the scatterer position in 3D), while providing five measurements, namely delay, AoA (azimuth and elevation), and AoD (azimuth and elevation).  Therefore, with enough resolvable NLOS paths, the UE can localize itself and also map the scatterers. The same holds for MIMO setups in the presence of an RIS with a blocked direct path, if the BS--RIS link is LOS, while the RIS--UE link includes resolvable NLOS components. In a general case, where both BS--RIS and RIS--UE links contain resolvable NLOS components, similar reasoning pertains only to RISs that are capable of baseband measurements collection at their side (e.g., as with the RIS design in~\cite{Alamzadeh2021ris}).

\section{An experimental localization case}\label{sec:experiment}

\begin{figure}
     \centering
     \begin{subfigure}{8cm}
         \centering
         \includegraphics[width=8cm]{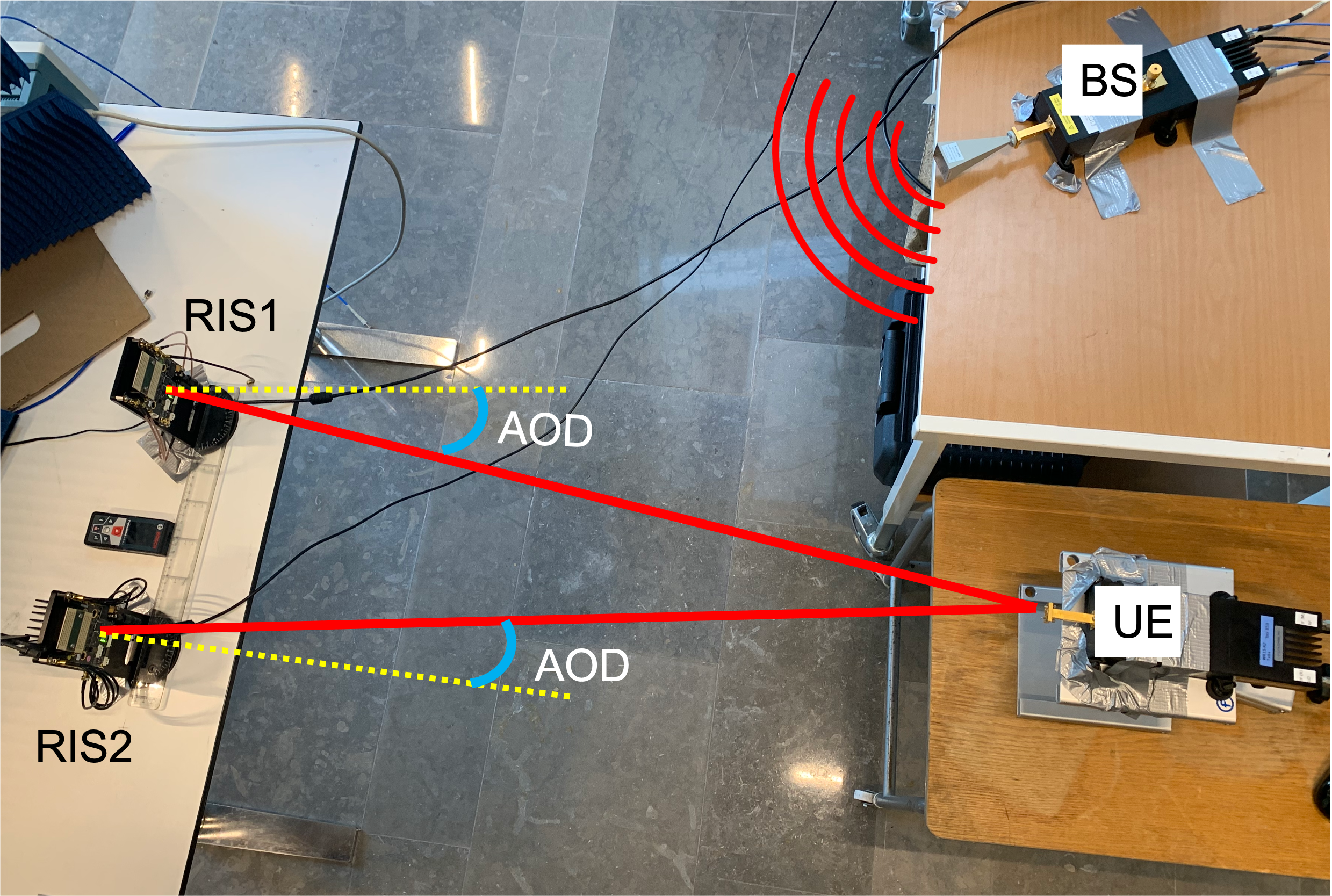}
         \caption{}
         \label{fig:Exp:system}
     \end{subfigure}
     \\
     \begin{subfigure}{8cm}
         \centering
         \begin{tikzpicture}
         \node(image) [anchor=south west] at (0,0)  {\includegraphics[width=8cm]{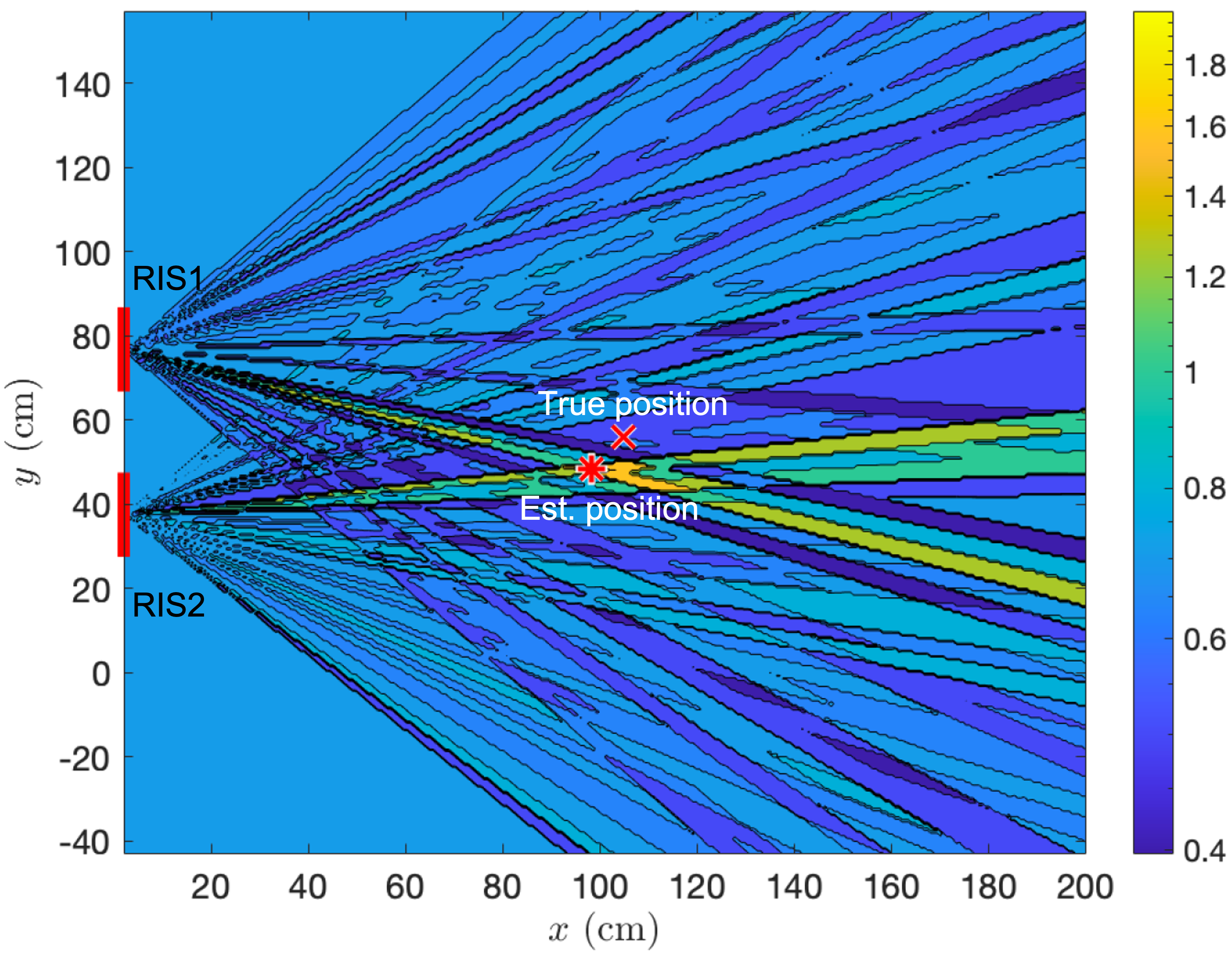}};
         \node(image) [fill=white,anchor=south west] at (3.4cm,-0cm)  {\footnotesize $x (\mathrm{cm})$};
         \node(image) [fill=white,rotate=90] at (0.15cm,3.4cm)  {\footnotesize $y (\mathrm{cm})$};
         \end{tikzpicture}
         
         \caption{}
         \label{fig:ExpResults}
     \end{subfigure}
        \caption{(a) The experimental localization setup at $60$ GHz and (b)  the normalized received power for different possible UE positions. The true and estimated UE positions are shown by a cross and a star, respectively.} 
        \label{fig:Experiments}
\end{figure}

In this section, we experimentally validate the SISO localization scenario with \emph{$1$ BS and $2$ RISs}, as described in Section\,\ref{sec:scenario:Siso2Ris1BS}. The laboratory experimental setup operating at $60$ GHz is illustrated in Fig.\,\ref{fig:Experiments}\,a. 
We deploy two commercial transceivers as RISs and a two-port millimeter wave vector network analyzer (VNA). We configure one of the VNA ports as the transmitter, i.e., the BS, and the other port as receiver, i.e., the UE. Two transceiver modules BFM6010 from SiversIMA, operating over the $57$--$71$ GHz frequency band, are used to emulate the two RISs. Each module contains a transmitter and a receiver, which are equipped with an antenna array of size 8$\times$3. To emulate an RIS with those modules, we loop the receiver I and Q signals back to the transmitter I and Q ports, and apply the desired beamforming (in terms of relative phase shifts) before transmitting the up-converted signal. A common local oscillator is used by the transmitter and the receiver for signal up- and down-conversion to avoid frequency offsets.

To perform localization in 2D, each of the two RISs iteratively sets $63$ beampattern configurations, which generate different azimuth \acp{aod}. The estimation of the \acp{aod} at the UE side is performed by selecting the beam that provides the largest received power for each RIS. Hence, the UE position is given by the intersection of such two beams. The sum of the normalized received powers from the two RISs (the values are normalized by the maximum received power from each RIS) corresponding to different \acp{aod} are depicted in Fig.\,\ref{fig:Experiments}\,(b). The maximum value in this figure represents the estimated UE position (marked by a red star), which is within the $10$-cm radius of the ground truth value (marked by a red cross).

\section{Conclusions and research challenges}
While reflective RISs are often put forward for improving localization accuracy via their capability of providing additional links with controlled SNR between BSs and UEs, another decisive breakthrough lies in their capacity to make localization feasible in lightweight operating contexts, where conventional systems would fail or necessitate much more resources. In this paper, after presenting the typical RIS operation and clarifying key concepts such as localization identifiability and ambiguity, we discussed and exemplified such benefits within a set of concrete canonical downlink localization scenarios, including one involving real pieces of hardware. Overall, the qualitative analysis of the presented scenarios confirms that the introduction of RISs can dramatically reduce the density, or even possibly, the hardware complexity of BSs (and thus accordingly, the overall network power consumption), which need to be deployed to ensure the identifiability of typical UE's state variables (e.g., position, velocity, and clock uncertainty). As a representative example of a single-antenna UE in the downlink, deploying two RISs and 1 single-antenna BS in a basic SISO setting is equivalent to more costly and power-greedy RIS-free settings, requiring for instance 2 multi-antenna BSs or even up to 4 single-antenna BSs. This equivalence is not only expected in terms of localization feasibility, but also even in terms of localization accuracy, given that power losses can be compensated by a sufficiently large number of integrated elements at the RISs.

To help this promising potential of RISs become a reality, the research community has been currently concentrating efforts to address open challenges related to RIS-enabled localization. Major research axes concern the realization of low-power and low-complexity RIS hardware devices with supportive element-wise reflection behavior (i.e., with respect to specific localization needs), the optimization of RIS deployment and placement on the field, as well as the design of localization-optimal RIS phase profiles, low-latency and low-overhead RIS control mechanisms enabling joint communication and localization/sensing services, and suitable and robust UE's state estimators capable of fully leveraging RIS capabilities (including the near-field regime).

\section*{Acknowledgment}
Authors would like to thank Hugo~Ssi~Yan~Kai and Viet~Lê for their contributions in the the experimental setup and  Gonzalo Seco Granados for the insightful discussions. This work was supported by the EU H2020 RISE-6G project under grant 101017011. 

\bibliographystyle{IEEEtran}
\bibliography{sample}

\newpage
\begin{IEEEbiographynophoto}
	{Kamran Keykhosravi} is a postdoctoral researcher with the Department of Electrical Engineering at Chalmers University of Technology, Sweden. His research interest include radio localization and reconfigurable intelligent surfaces.
  \end{IEEEbiographynophoto}
  \vspace{-1.2cm}
\begin{IEEEbiographynophoto}
	{Benoît Denis} is a senior researcher with the Department of Wireless Technologies at CEA-Leti, Grenoble, France. His main research interests concern wireless localization and location-enabled networks, with a focus on cooperative, multipath-aided and data fusion approaches, as well as on enabling technologies such as ultra wideband radio, millimeter waves or reconfigurable intelligent surfaces.
  \end{IEEEbiographynophoto}
  \vspace{-1.2cm}

\begin{IEEEbiographynophoto}
{George C. Alexandropoulos} 
is Assistant Professor for wireless communication systems and signal processing with the Department of Informatics and Telecommunications, National and Kapodistrian University of Athens, Greece. His research interests span the general areas of algorithmic design and performance analysis for wireless networks with emphasis on multi-antenna transceiver hardware architectures, active and passive metasurfaces, full duplex radios, and high-frequency communications, as well as distributed machine learning algorithms. 
  \end{IEEEbiographynophoto}
  \vspace{-1.2cm}

\begin{IEEEbiographynophoto}
	{Zhongxia Simon He} 
	received the M.Sc. degree from the Beijing Institute of Technology, Beijing, China, and the Ph.D. degree from the Chalmers University of Technology, Gothenburg, Sweden, in 2008 and 2014, respectively. He is currently an Associate Professor with the Microwave Electronics Laboratory, Department of Microtechnology and Nanoscience (MC2), Chalmers University. His current research interests include high data rate wireless communication, modulation and demodulation, mixed-signal integrated circuit design, radar, and packaging. He is also jointly working at SinoWave AB, Sweden.
	
  \end{IEEEbiographynophoto}
  \vspace{-1.2cm}

\begin{IEEEbiographynophoto}
	{Antonio Albanese} 
	received the M.Sc. degree in Telecommunications Engineering from Politecnico di Milano, Italy, in 2018. Currently, he is pursuing his Ph.D. in Telematic Engineering at Universidad Carlos III de Madrid, Spain, while being appointed as Research Scientist at NEC Laboratories Europe GmbH, Heidelberg, Germany. His research field covers millimeter waves, reconfigurable intelligent surfaces, applied mathematical optimization and machine learning techniques, with a particular interest in localization and prototyping. 
  \end{IEEEbiographynophoto}
  \vspace{-1.2cm}

\begin{IEEEbiographynophoto}
{Vincenzo Sciancalepore} 
received his M.Sc. degree in Telecommunications Engineering and Telematics Engineering in 2011 and 2012, respectively, whereas in 2015, he received a double Ph.D. degree. Currently, he is a Principal Researcher at NEC Laboratories Europe in Heidelberg, focusing his activity on network virtualization and network slicing challenges. He is the Chair of the ComSoc Emerging Technologies Initiative (ETI) on Reconfigurable Intelligent Surfaces (RIS) and an editor of IEEE Transactions on Wireless Communications.
 \end{IEEEbiographynophoto}
  \vspace{-1.2cm}

\begin{IEEEbiographynophoto}
	{Henk Wymeersch} is Professor at Chalmers University of Technology, Sweden, active in the area of 5G and beyond 5G radio localization and sensing. 
  \end{IEEEbiographynophoto}

\if{0}
\begin{table*}
\centering
\caption{Identifiability analysis of DL localization (not synchronization or
orientation estimation). \textbf{maybe a series of small figures would be better?}}
\resizebox{0.99\textwidth}{!} {
\begin{tabular}{|c|c|c|c|c|c|}
\hline 
 & \textbf{user state} & \textbf{Measurements} & \textbf{\#BS needed w/o RIS} & \textbf{\#BS needed with 1 RIS} & \textbf{no LOS to BS}\tabularnewline
\hline 
\hline 
\textbf{SISO-DL} & pos, clk & delay & $4$ TDoAs or $3$ RTTs & $1$ BS & yes, with near-field\tabularnewline
\hline 
\textbf{MISO-DL} & pos, clock & delay, AoD & $2$ TDoAs + AoD or $1$ RTT + AoD & $1$ BS (also narrowband) & yes, with near-field\tabularnewline
\hline 
\textbf{SIMO-DL} & pos, ori, clock & delay, AoA & $4$ TDoAs or $3$ RTTs & $1$ BS & yes, with near-field\tabularnewline
\hline 
\textbf{MIMO-DL} & pos, ori, clock & delay, AoA, AoD & $2$ TDoAs + AoD or $1$ RTT + AoD & $1$ BS (also narrowband) & yes, with near-field\tabularnewline
\hline 
\textbf{SISO-UL} & pos, clock & delay & $4$ TDoAs or $3$ RTTs & $0$ BS in FD, $1$ BS in HD & N/A\tabularnewline
\hline 
\textbf{SIMO-UL} & pos, clock & delay, AoD & $2$ TDoAs + AoA or $1$ RTTs + AoA & $0$ BS in FD, $1$ BS in HD & yes, with near-field, HD\tabularnewline
\hline 
\textbf{MISO-UL} & pos, ori, clock & delay, AoA & $4$ TDoAs or $3$ RTTs & $0$ BS in FD, $1$ BS in HD & yes, with near-field, HD\tabularnewline
\hline 
\textbf{MIMO-UL} & pos, ori, clock & delay, AoA, AoD & $2$ TDoAs + AoA or $1$ RTTs + AoA & $0$ BS in FD, $1$ BS in HD & yes, with near-field, HD\tabularnewline
\hline 
\end{tabular}
}
\end{table*}
\fi
\end{document}

%% file: acs.tex
\acrodef{bs}[BS]{base station}
\acrodef{ue}[UE]{user equipment}
\acrodef{los}[LOS]{line-of-sight}
\acrodef{aoa}[AoA]{angle-of-arrival}
\acrodef{aod}[AoD]{angle-of-departure}
\acrodef{doa}[DoA]{direction-of-arrival}
\acrodef{dod}[DoD]{direction-of-departure}
\acrodef{toa}[ToA]{time-of-arrival}
\acrodef{tdoa}[TDoA]{time-difference-of-arrival}
\acrodef{ris}[RIS]{reconfigurable intelligent surface}
\acrodefplural{ris}[RISs]{reconfigurable intelligent surfaces}
\acrodef{tx}[Tx]{transmitter}
\acrodef{rx}[Rx]{receiver}
\acrodefplural{rx}[Rxs]{receivers}
\acrodef{crb}[CRB]{Cram\'er-Rao lower bounds}
\acrodef{rss}[RSS]{received signal strength}
\acrodef{rtt}[RTT]{round-trip time}
\acrodef{los}[LOS]{line-of-sight}
\acrodef{nlos}[NLOS]{non line-of-sight}
\acrodef{dft}[DFT]{discrete Fourier transform}
\acrodef{fft}[FFT]{fast Fourier transform}
\acrodef{fim}[FIM]{Fisher information matrix}
\acrodef{upa}[UPA]{uniform planar array}
\acrodefplural{upa}[UPAs]{uniform planar arrays}
\acrodef{peb}[PEB]{position error bound}
\acrodef{snr}[SNR]{signal-to-noise ratio}
\acrodef{sre}[SRE]{smart radio environment}
\acrodefplural{sre}[SRE]{smart radio environments}
\acrodef{mimo}[MIMO]{multiple-input  multiple-output}
\acrodef{rfid}[RFID]{radio-frequency identification}
\acrodef{siso}[SISO]{single-input single-output}
\acrodef{simo}[SIMO]{single-input multiple-output}
\acrodef{ici}[ICI]{inter-carrier interference}
\acrodef{iid}[iid]{independent and identically distributed}
\acrodef{qos}[QoS]{quality of service}
\acrodef{miso}[MISO]{multiple-input  single-output}
\acrodef{gdop}[GDoP]{Geometric Dilution of Precision}
\acrodef{ofdm}[OFDM]{orthogonal frequency-division multiplexing}
\acrodef{uav}[UAV]{unmanned aerial vehicle}
\acrodef{nb}[NB]{narrow-band}
\acrodef{wb}[WB]{wide-band}